\documentclass[%
reprint,
amsmath,amssymb,
aps,
prl
]{revtex4-2}

\usepackage{graphicx}% Include figure files
\usepackage{dcolumn}% Align table columns on decimal point
\usepackage{bm}% bold math
\begin{document}
\preprint{APS/123-QED}

\title{$\mathcal{PT}$-symmetry-breaking enhanced cavity optomechanical magnetometry}
\author{Zhucheng Zhang,$^{1}$ Yi-Ping Wang,$^{2}$ and Xiaoguang Wang$^{1,3}$}
\email{xgwang1208@zju.edu.cn}
\affiliation{$^{1}$Zhejiang Institute of Modern Physics, Department of Physics, Zhejiang University, Hangzhou 310027, China\\
$^{2}$College of Science, Northwest A$\&$F University, Yangling 712100, China\\
$^{3}$Graduate School of China Academy of Engineering Physics, Beijing 100193, China}
\date{\today}

\begin{abstract}
$\mathcal{PT}$-symmetry-breaking enhanced cavity optomechanical magnetometer is proposed, which is achieved by monitoring the change of intensity of
a nonlinear four-wave mixing (FWM) process in a gain-cavity-assisted cavity optomechanical system (COMS). Compared with the traditional single loss COMS, the FWM intensity can be enhanced by two orders of magnitude when the gain-cavity-assisted COMS operates at the $\mathcal{PT}$-symmetry-breaking phase.
Meanwhile, the sensitivity of magnetic field sensing can be increased from $10^{-9}$T to $10^{-11}$T. This originally comes from the fact that the
effective detuning and decay of loss-cavity can be effectively modified in the $\mathcal{PT}$-symmetry-breaking phase. Our work shows that
an ultrahigh-sensitivity magnetometer can be achieved in the $\mathcal{PT}$-symmetry-breaking COMS, which will have wide applications in the field
of quantum sensing.
\end{abstract}
\maketitle

$\textit{Introduction}.$---Ultrahigh-sensitivity magnetometers with small size play an important role in medicine, geology, biology, defense and so
on~\cite{edelstein2007advances,pham2011magnetic,bouchard2011detection,hall2010monitoring},
which attracts great interest of researchers. Although the sensitivity of magnetometers based on atom and magnetostrictive material can achieve
a magnitude of aT\,Hz$^{-1/2}$ and fT\,Hz$^{-1/2}$, the size scales of these systems are generally limited to millimeter or 
centimeter~\cite{dang2010ultrahigh,zhai2006detection}. Besides,
magnetometers based on superconducting quantum interference device and NV center are hampered by operating temperature, fabrication
issues and readout schemes, respectively~\cite{romalis2011atomic,balasubramanian2009ultralong,schoenfeld2011real,pham2011magnetic}. To improve the sensitivity and reduce the size of magnetometer is still the focus of designing systems.

Cavity optomechanics is a hot research field exploring the nonlinear interaction between electromagnetic radiation and nano- and
micro-mechanical systems~\cite{aspelmeyer2014cavity}, which provides a promising platform for many theoretical and experimental researches, such as quantum ground-state cooling of mechanical oscillators~\cite{poggio2007feedback,bhattacharya2007trapping,schliesser2008resolved}, optomechanically induced transparency~\cite{weis2010optomechanically,agarwal2010electromagnetically,kronwald2013optomechanically}, normal-mode 
splitting~\cite{dobrindt2008parametric,groblacher2009observation,zhang2019normal} and so on. With this nonlinear optomechanical interaction, a
micron-scale cavity optomechanical magnetometer with room temperature operation has realized a peak magnetic field sensitivity of 400nT\,Hz$^{-1/2}$ via
the magnetic-field-induced deformations of a magnetostrictive material in experiment~\cite{forstner2012cavity}. Besides, it has been shown that thanks to this nonlinear interaction, the lower and upper motional sidebands can be generated in the transmission spectra of cavity optomechanical system 
(COMS)~\cite{schliesser2008resolved,xiong2012higher}. Based on these motional sidebands,
the weak magnetic field can also be precisely measured by finding out the correlations between the structure of transmission spectra and the measured
magnetic field. For example, through monitoring the deformation of optomechanically induced transparency window (corresponding to the upper first order
sideband)~\cite{liu2017proposed}, or the change of intensity of the upper second order sideband~\cite{liu2018precision}, cavity optomechanical magnetometers can achieve a sensitivity of nT through electromagnetic interactions in theories. Furthermore, utilizing these motional sidebands, COMS can also be used to sense other physical quantities, such as electrical charges~\cite{zhang2012precision,li2018enhanced}, environmental temperature~\cite{wang2015precision}, mass~\cite{he2015sensitivity,bin2019mass} and so on. For the COMS as an all-optical sensor, these motional sidebands are
undoubtedly a powerful tool. 

On the other hand, since the concept of parity-time ($\mathcal{PT}$)-symmetry was put forward, it has been widely studied in theories and 
experiments~\cite{konotop2016nonlinear}, such as $\mathcal{PT}$-symmetric phonon laser~\cite{jing2014pt}, $\mathcal{PT}$-symmetry-breaking 
chaos~\cite{lu2015p}, loss-induced transparency~\cite{guo2009observation}, low-power optical diodes~\cite{peng2014loss} and a single-mode 
laser~\cite{hodaei2014parity,feng2014single}. With the singular characteristics of the $\mathcal{PT}$-symmetric system operating at the phase transition 
from unbroken to broken $\mathcal{PT}$-symmetry, $\mathcal{PT}$-symmetric system can also be used as sensors for 
particle~\cite{wiersig2014enhancing,chen2017exceptional}, acoustics~\cite{fleury2015invisible} and mechanical motion~\cite{liu2016metrology}. Besides, the concept of $\mathcal{PT}$-symmetry was introduced into the quantum noise theory to calculate the
signal-to-noise performance~\cite{zhang2019quantum}. Recently, the generation of motional sidebands has been shown to be enhanced in the
$\mathcal{PT}$-symmetric COMS~\cite{he2019parity}. Thus, a natural question is whether the $\mathcal{PT}$-symmetric COMS combined with motional
sidebands can enhance the sensitivity of magnetic field sensing significantly, which will be a significant improvement for the magnetometer based on COMS.

\begin{figure*}
	\centering
	\includegraphics[width=0.95\linewidth]{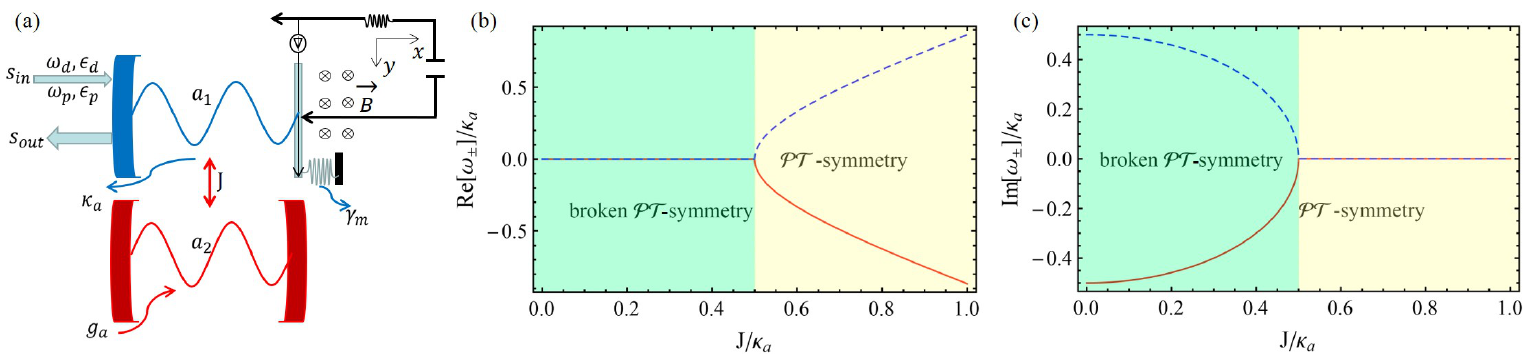}
	\caption{(color online). (a) Schematic diagram of a $\mathcal{PT}$-symmetric COMS, in which a loss COMS [with decay rate $\kappa_a$ ($\gamma_m$) of cavity (mechanical) mode] is coupled to a gain-cavity (with optical gain rate $g_a$ of cavity mode) through a tunneling coupling $J$. Besides, the loss COMS is driven by a strong driving field of frequency $\omega_d$ with amplitude $\epsilon_d$ and probed by a weak probe field of frequency
		$\omega_p$ with amplitude $\epsilon_p$, meanwhile, the movable end (as an electromechanical oscillator) is passed through a surface current with intensity $I$ and the measured	magnetic field with strength $B$ is applied to the loss COMS. (b) and (c) Real and imaginary parts of eigenfrequencies
		$\omega_\pm$ obtained by diagonalizing the Hamiltonian of $\mathcal{PT}$-symmetric COMS under the weak-coupling regime \cite{graefe2010quantum,ramezani2010unidirectional}. The gain-cavity-assisted COMS with balanced gain $g_a$ and loss $\kappa_{a}$ can show a phase transition from broken to unbroken $\mathcal{PT}$-symmetry by controlling the tunneling coupling.}
\end{figure*}
In this Letter, we propose a $\mathcal{PT}$-symmetry-breaking enhanced cavity optomechanical magnetometer by monitoring the change of intensity of the
lower first order sideband in a gain-cavity-assisted cavity optomechanical system, which corresponds to a nonlinear four-wave mixing (FWM)
process~\cite{kippenberg2004kerr}. Compared with the traditional single loss COMS, we show that the FWM intensity in the $\mathcal{PT}$-symmetric COMS
can be enhanced by two orders of magnitude. What's more, the measurement precision of weak magnetic field based on the change of FWM intensity can be
increased from 10$^{-9}$T to 10$^{-11}$T. Physically, when the $\mathcal{PT}$-symmetry-breaking phase occurs, the optical gain of gain-cavity can completely
balance the effective decay of the loss-cavity; besides, the effective loss-cavity detuning can also tend to zero, i.e., the loss-cavity is driven almost resonantly.
This ultimately leads to the result that the sensitivity of the $\mathcal{PT}$-symmetric COMS can be higher than the traditional single loss COMS. This
unconventional optomechanical magnetometer combines $\mathcal{PT}$-symmetry and motional sideband, which will have wide applications in the field of
precision measurement.

$\textit{Model.}$---The proposed $\mathcal{PT}$-symmetric COMS is shown in Fig.~1(a), in which the measured weak magnetic field is applied to our system through the electromagnetic interaction. The Hamiltonian of system can be written as
\begin{align}
	H&=-\hbar\Delta_{a}(a_{1}^{\dagger}a_{1}+a_{2}^{\dagger}a_{2})+\frac{p^{2}}{2m}+\frac{1}{2}m\omega_{m}^{2}x^{2} \notag \\
	&+\hbar J(a_{1}^{\dagger}a_{2}+a_{2}^{\dagger}a_{1})+\hbar Ga_{1}^{\dagger}a_{1}x+\zeta Bx \notag \\
	&+i\hbar\sqrt{\eta_{c}\kappa_{a}}[(\epsilon_{d}a_{1}^{\dagger}+\epsilon_{p}a_{1}^{\dagger}e^{-i\Omega t})-h.c.],
\end{align}
in which $a_1$ ($a_{1}^\dagger$) and $a_2$ ($a_{2}^\dagger$) are the annihilation (creation) operators of loss- and gain-cavity, respectively.
$\Delta_{a}=\omega_{d}-\omega_a$ and $\Omega=\omega_{p}-\omega_d$ are the frequency detunings with the cavity resonant frequency $\omega_a$. $p$ ($x$)
describes the momentum (position) operator of the oscillator (with resonance frequency $\omega_m$ and mass $m$). The term $\hbar Ga_{1}^{\dagger}a_{1}x$
characterizes the optomechanical interaction between the loss-cavity and the oscillator with coupling strength $G$. The parameter $\zeta$ represents the
strength of electromagnetic interaction, which is proportional to the current intensity $I$ and the effective range of action \cite{liu2017proposed,liu2018precision}. 
The last term describes the interactions
between the input fields and the loss-cavity with a critical coupling parameter $\eta_{c}=1/2$. $\epsilon_{d,p}=\sqrt{P_{d,p}/\hbar \omega_{d,p}}$ is the amplitude of the input fields with the power $P_{d,p}$, and $\kappa_{a}$ is the total decay rate of the loss-cavity. As shown in Fig.~1(b) and Fig.~1(c), when the
tunneling coupling $J=0.5\kappa_{a}$, the gain-cavity-assisted COMS with balanced gain $g_a$ and loss $\kappa_{a}$ can show a $\mathcal{PT}$-symmetry phase transition, in
which the eigenfrequencies and the corresponding eigenstates of system coalesce simultaneously \cite{konotop2016nonlinear}. One can find in the following
sections that the system can show a better sensitivity for the change of weak magnetic field at this $\mathcal{PT}$-symmetry phase transition point.

\begin{figure*}
	\centering
	\includegraphics[width=0.325\linewidth]{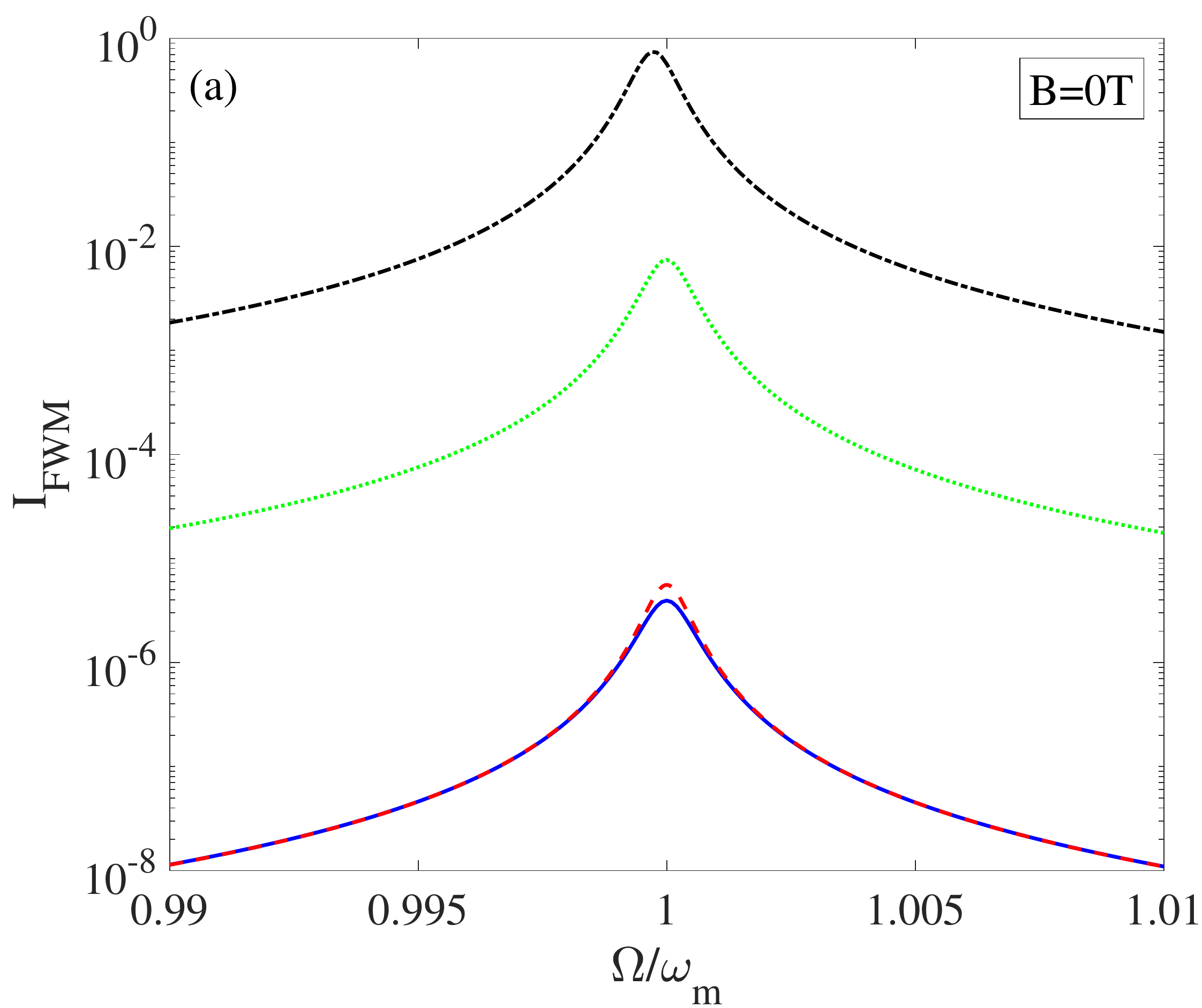}
	\includegraphics[width=0.325\linewidth]{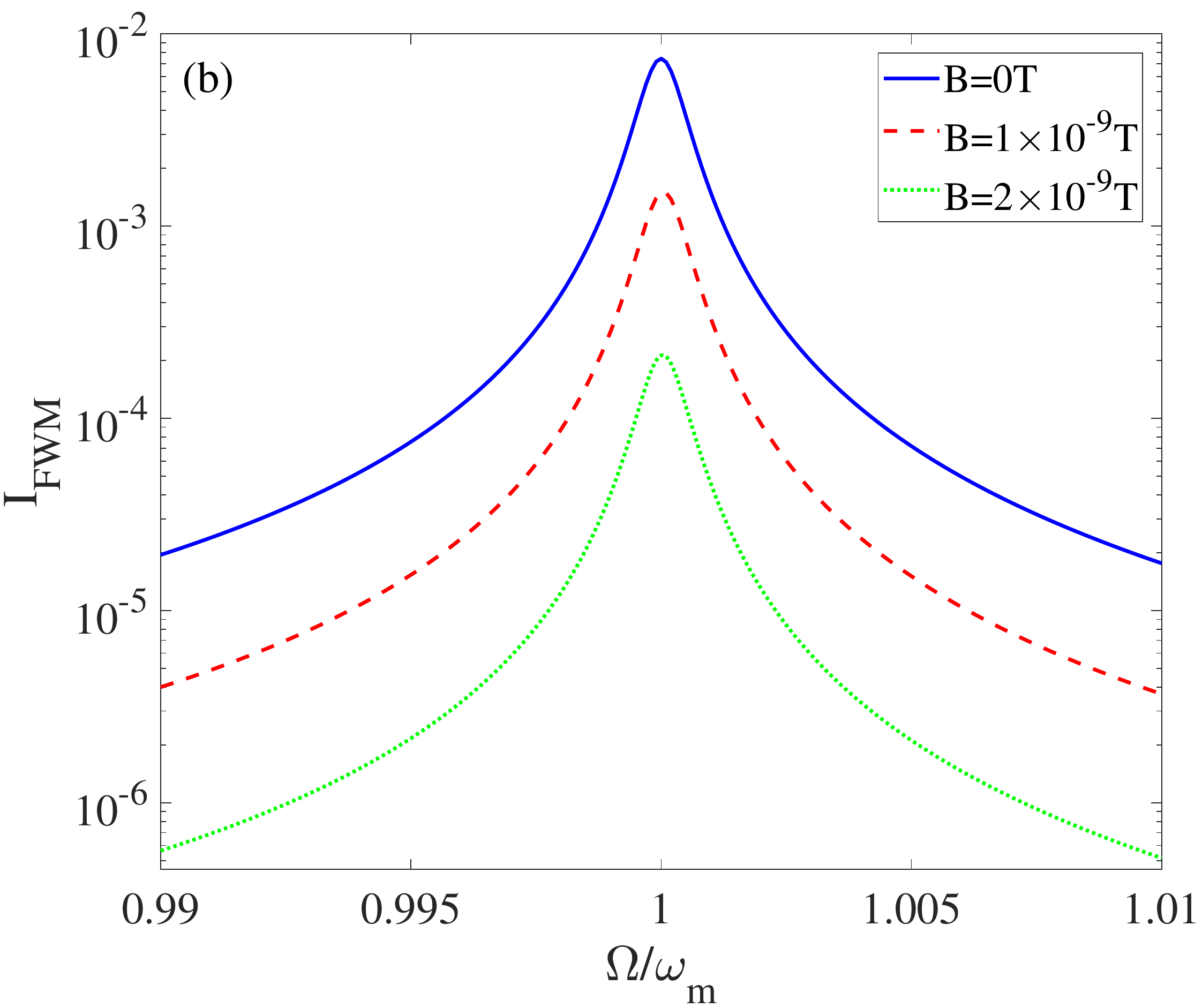}
	\includegraphics[width=0.325\linewidth]{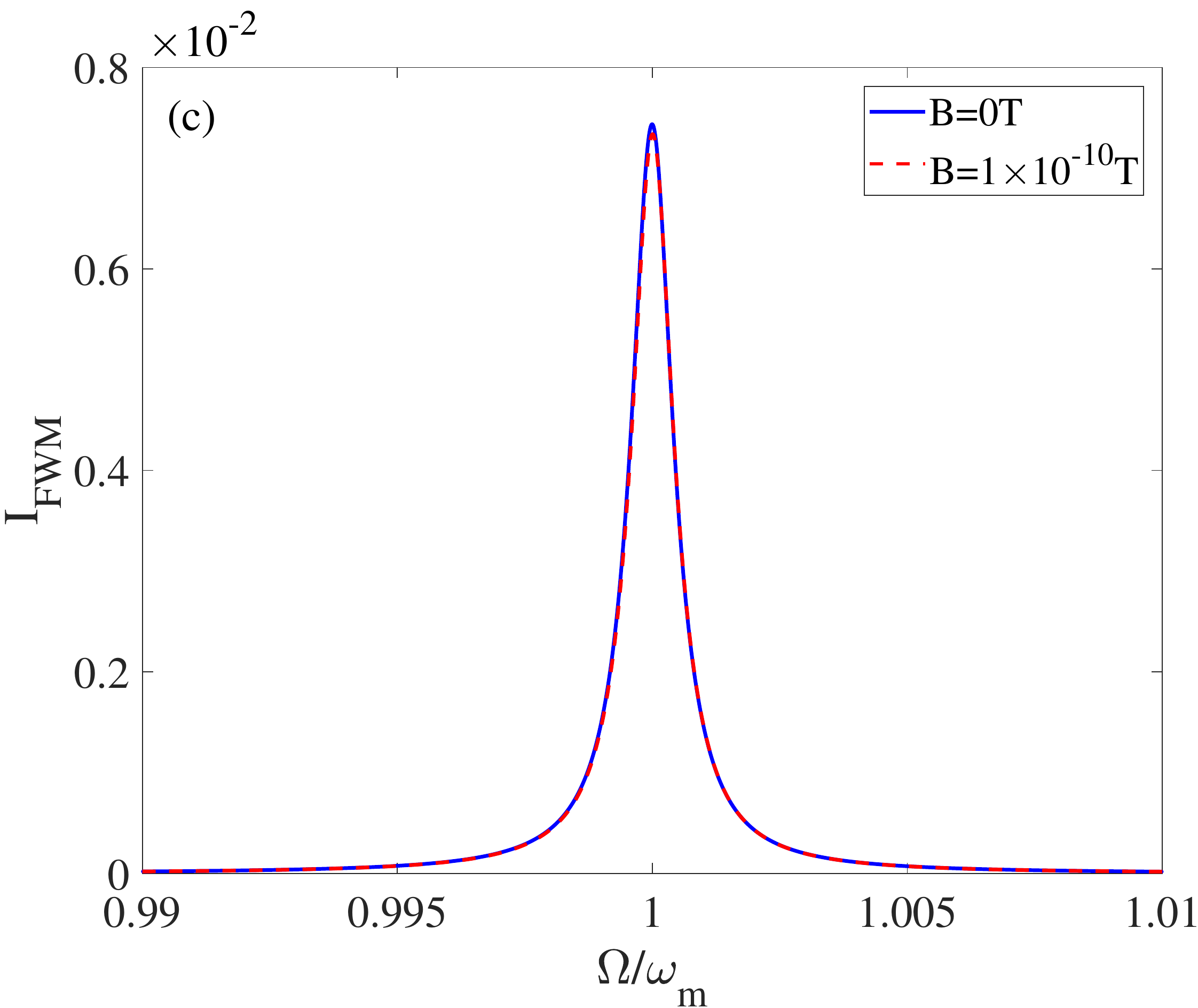}
	\caption{FWM intensity spectrum as a function of detuning $\Omega$. (a) The solid blue and dashed red curves represent
		the loss-cavity is driven by blue-detuning (i.e., $\Delta_{a}=\omega_m$) and red-detuning (i.e., $\Delta_{a}=-\omega_m$), respectively, and the
		tunneling coupling $J=0$; the dotted green and dot-dashed black curves represent the tunneling coupling $J=0$ and $J=0.5\kappa_{a}$,
		respectively, and the loss-cavity is driven resonantly, i.e., $\Delta_a=0$. (b) and (c) The loss-cavity is driven resonantly, and the tunneling
		coupling $J=0$. The parameters are: $\omega_m=2 \pi \times 0.1\textrm{MHz},~m=100\textrm{pg},~G=-2\pi \times 11\textrm{MHz/nm},~\gamma_{m}=2\pi \times 0.1\textrm{kHz},~\kappa_{a}=0.1\omega_m,~g_{a}=\kappa_{a},~\zeta=2\times10^{-5}\textrm{A}\cdot \textrm{m}$ for the current intensity $I = 1 \textrm{mA},~P_{d}=1\textrm{pW}$, and the wavelength of driving field $\lambda_d=2\pi c/\omega_{d}=532\textrm{nm}$ ($c$ represents the speed of light in vacuum).}
\end{figure*}
With the semiclassical Langevin equations [i.e., setting $o(t)\equiv\left\langle o(t)\right\rangle$, $o=a_{1,2},~x,~p$], the dynamics evolution of system can be
described by the following equations,
\begin{align}
\frac{da_{1}}{dt}&=(i\Delta_{a}-iGx-\frac{\kappa_{a}}{2})a_{1}-iJa_{2} \notag \\
&+\sqrt{\eta_{c}\kappa_{a}}(\epsilon_{d}+\epsilon_{p}e^{-i\Omega t}),\\
\frac{da_{2}}{dt}&=(i\Delta_{a}+\frac{g_{a}}{2})a_{2}-iJa_{1},\\
\frac{d^{2}x}{dt^{2}}&=-\gamma_{m}\frac{dx}{dt}-\omega_{m}^{2}x-\frac{1}{m}(\hbar Ga_{1}^{\dagger}a_{1}+\zeta B).
\end{align}
Due to the fact that the probe field is much weaker than the driving field, the above dynamic equations can be solved with the perturbation method. Using
$o=o_s+\delta o$ with $o_s$ and $\delta o$ being the steady-state values and the corresponding perturbation terms, respectively, one can get the following
steady-state values,
\begin{align}
a_{1s}&=-\frac{\sqrt{\eta_{c}\kappa_{a}}\epsilon_{d}}{i\Delta_{\textrm{eff}}-\frac{\kappa_{\textrm{eff}}}{2}}, \\
a_{2s}&=\text{\ensuremath{\frac{iJa_{1s}}{i\Delta_{a}+\frac{g_{a}}{2}}}},\\
x_{s}&=-\frac{\hbar GN_{1}+B\zeta}{m\omega_{m}^{2}},
\end{align}
with the effective detuning $\Delta_{\textrm{eff}}$ and decay rate $\kappa_{\textrm{eff}}$ of loss-cavity due to the three interactions in our system, i.e., optomechanical, 
electromagnetic and double-cavity tunneling interactions,
\begin{align}
\Delta_{\textrm{eff}}&=\Delta_{a}-[Gx_{s}+J^{2}\Delta_{a}/(\Delta_{a}^{2}+g_{a}^{2}/4)],\\
\kappa_{\textrm{eff}}&=\kappa_{a}-J^{2}g_{a}/(\Delta_{a}^{2}+g_{a}^{2}/4),
\end{align}
and the average photon number of loss-cavity $N_1=\left|a_{1s}\right| ^{2}$. One can find that the effective detuning $\Delta_{\textrm{eff}}$ and the effective
decay rate $\kappa_{\textrm{eff}}$ are related to the driving detuning $\Delta_{a}$ and the tunneling coupling $J$.

Besides, the corresponding evolution of the perturbation terms can be derived as,
\begin{align}
\frac{d\delta a_{1}}{dt}&=(i\Delta-\frac{\kappa_{a}}{2})\delta a_{1}-iJ\delta a_{2} \notag \\
&-iG(a_{1s}\delta x+\delta a_{1}\delta x)+\sqrt{\eta_{c}\kappa_{a}}\epsilon_{p}e^{-i\Omega t},\\
\frac{d\delta a_{2}}{dt}&=(i\Delta_{a}+\frac{g_{a}}{2})\delta a_{2}-iJ\delta a_{1},\\
\frac{d^{2}\delta x}{dt^{2}}&=-\gamma_{m}\frac{d\delta x}{dt}-\omega_{m}^{2}\delta x \notag \\
&-\frac{\hbar G}{m}(a_{1s}\delta a_{1}^{*}+a_{1s}^{*}\delta a_{1}+\delta a_{1}^{*}\delta a_{1}),
\end{align}
in which $\Delta=\Delta_{a}-Gx_s$.
In order to solve the above nonlinear equations, we make the following ansatz:
\begin{align}
\delta a_{1}&=A_{1}^{u}e^{-i\Omega t}+A_{1}^{l}e^{i\Omega t},\\
\delta a_{2}&=A_{2}^{u}e^{-i\Omega t}+A_{2}^{l}e^{i\Omega t},\\
\delta x&=X_{1}e^{-i\Omega t}+X_{1}^{*}e^{i\Omega t},
\end{align}
in which $A_{1}^{l}$ ($A_{1}^{u}$) represents the coefficient of the lower (upper) first order sideband of loss COMS, which corresponds to the
nonlinear four-wave mixing (FWM) process ~\cite{schliesser2008resolved,xiong2012higher,kippenberg2004kerr}.
We should note that our scheme focuses on the combined effects of the $\mathcal{PT}$-symmetry and the lower first order sideband on the weak magnetic field sensing. By substituting the ansatz into Eqs.~(10)-(12), we can get the solution for
$A_{1}^{l}$,
\begin{equation}
A_{1}^{l}=\frac{ia_{1s}^{2}\epsilon_{p}G^{2}\hbar\sqrt{\eta_{c}\kappa_{a}}}{D(\Omega)},
\end{equation}
with
\begin{align}
D(\Omega)&=2\left|a_{1s}\right|^{2}\!G^{2}\hbar D_{1}(\Omega)\!+\!mD_{2}(\Omega)\!D_{3}(\Omega)\!D_{4}(\Omega),\\
D_{1}(\Omega)&=\Delta-\frac{\Delta_{a}J^{2}}{\Delta_{a}^{2}-(i\frac{g_{a}}{2}+\Omega)^{2}},\\
D_{2}(\Omega)&=[\Delta+(i\frac{\kappa_{a}}{2}-\Omega)]-\frac{J^{2}}{\Delta_{a}-(i\frac{g_{a}}{2}+\Omega)},\\
D_{3}(\Omega)&=[\Delta-(i\frac{\kappa_{a}}{2}-\Omega)]-\frac{J^{2}}{\Delta_{a}+(i\frac{g_{a}}{2}+\Omega)},\\
D_{4}(\Omega)&=\omega_{m}^{2}-\Omega^{2}+i\gamma_{m}\Omega.
\end{align}
Then, according to the input-output relation \cite{gardiner2004quantum}, we can define the intensity of FWM in terms of the probe field,
\begin{equation}
\textrm{I}_{\textrm{FWM}}=\left|\frac{\sqrt{\eta_{c}\kappa_{a}}A_{1}^{l}}{\epsilon_{p}}\right|^{2}.
\end{equation}

$\textit{Magnetic\! field\! sensing\! with\! single\! loss\! COMS.}$---At first, we study the performance of magnetic field sensing based on the change of
FWM intensity with single loss COMS. The parameters used in our numerical simulations are analogous to the related theoretical and
experimental works \cite{thompson2008strong,liu2017proposed,liu2018precision}. 
As shown in Fig.~2, the FWM intensity spectrum is plotted as a function of the detuning $\Omega$. In our scheme, the loss COMS
is driven resonantly, which can significantly enhance the intensity of FWM compared to the red- and blue-detuning, as shown in Fig.~2(a). With
the resonantly driven single loss COMS, one can find that the intensity of FWM decreases with the increase of the measured weak magnetic field 
[see Fig.~2(b)]. Meanwhile, the sensitivity of this system can distinguish the magnetic field with only about 10$^{-9}$T [see Fig.~2(c)]. Physically,
for the resonantly driven single loss COMS, the effective detuning $\Delta_{\textrm{eff}}$ of loss-cavity is simplified as $-Gx_s$. Due to the presence of optomechanical
interaction, the value of the effective detuning isn't zero [see the inset of Fig.~3(a)], but with the increase of the measured magnetic field, it
quickly decreases to zero first, then its absolute value increases gradually, which shows that the driving of the single loss COMS deviates from
the resonance condition gradually even if there is a small region close to resonance. Besides, from Fig.~3(b), one can find that due to the small
region close to resonance of the effective detuning, the average photon number of loss-cavity slightly increases first, but it also gradually
decreases, which is consistent with the change trend of the effective detuning. This ultimately leads to the result that the FWM intensity generated in the single
loss COMS decreases with the increase of the measured magnetic field, which also limits the performance of system for magnetic field sensing.
\begin{figure}
	\centering
	\includegraphics[width=0.8\linewidth]{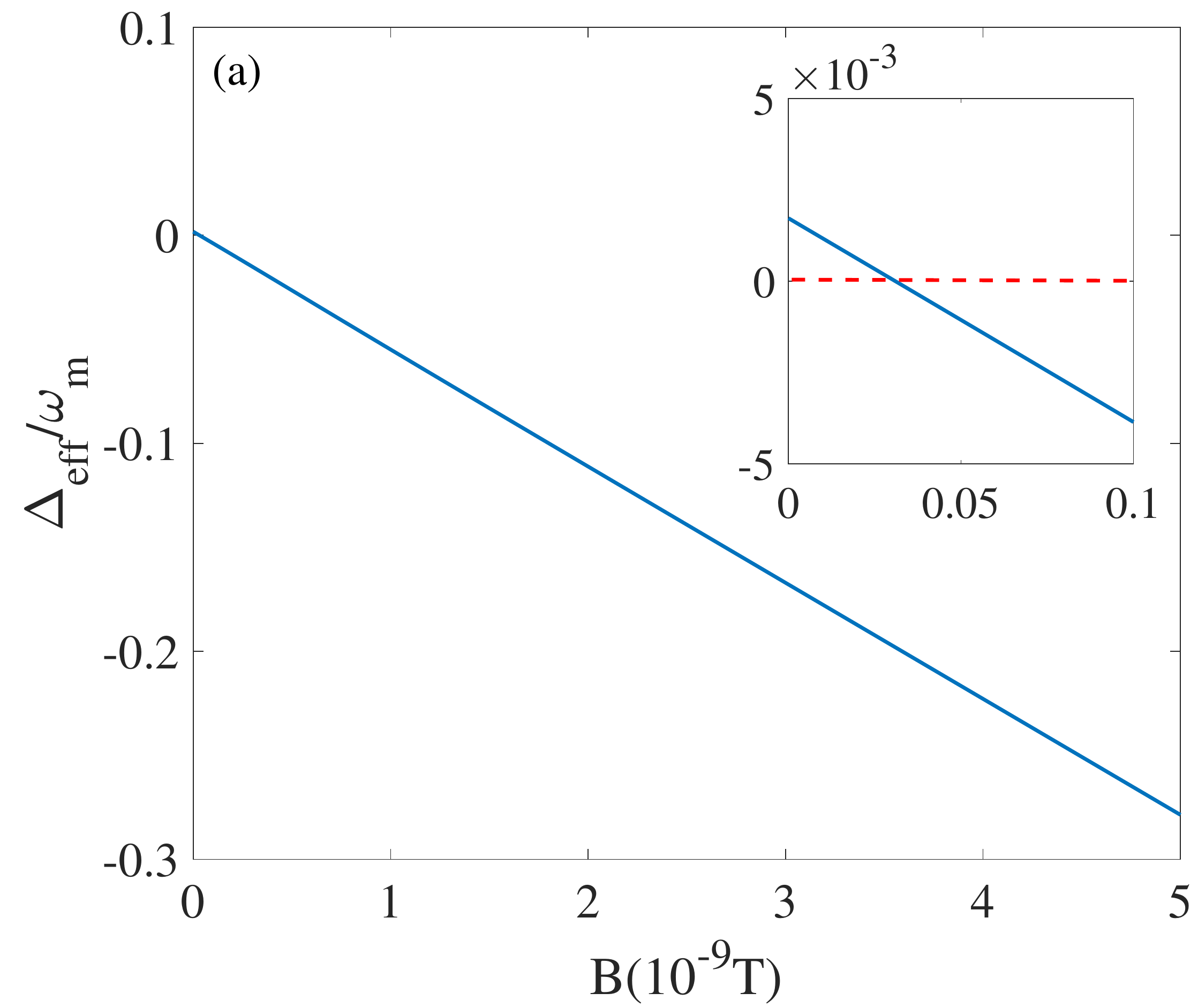}\\
	\includegraphics[width=0.8\linewidth]{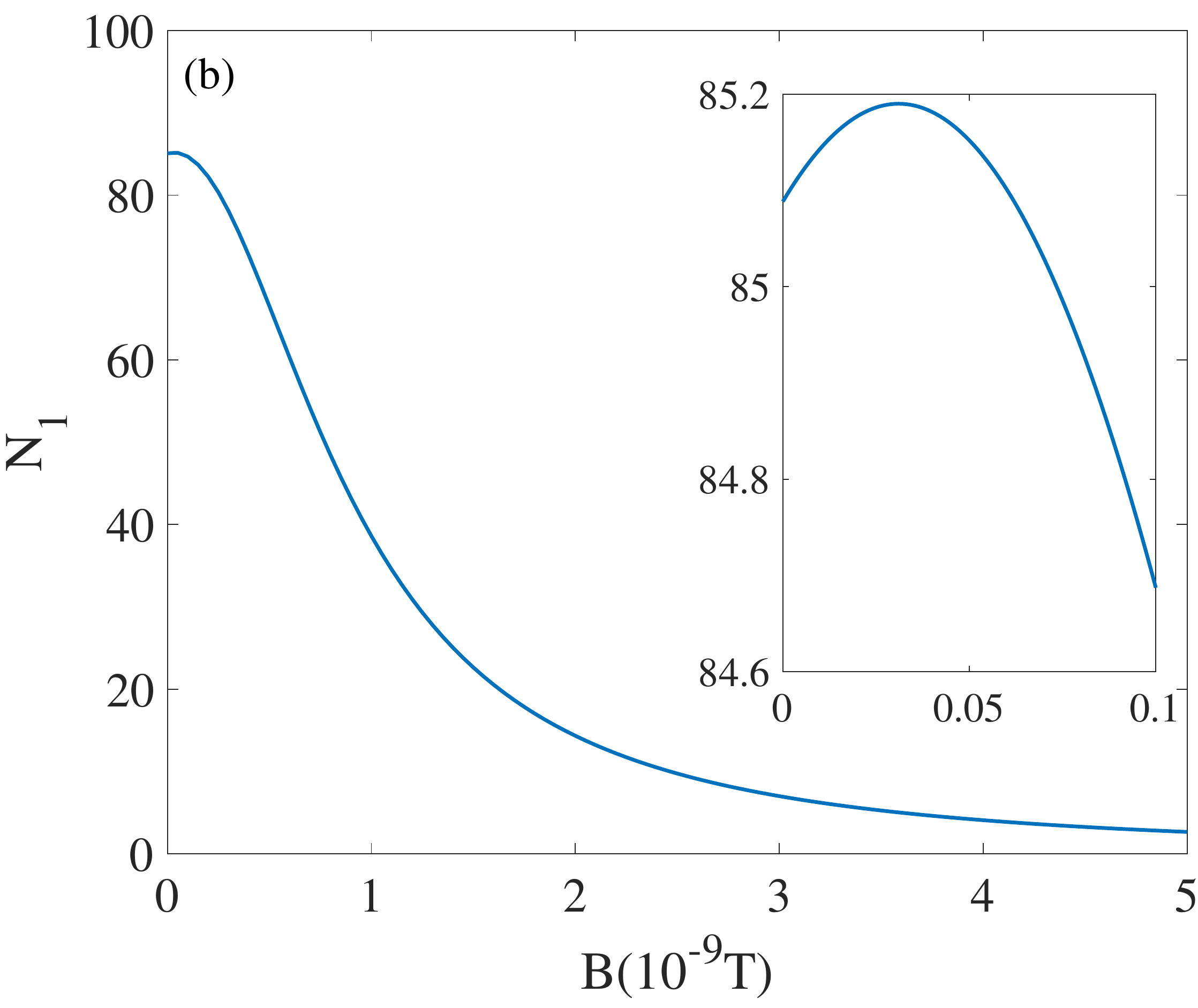}
	\caption{(a) Effective detuning $\Delta_{\textrm{eff}}$ and (b) average photon number $N_1$ of loss-cavity as a function of
		strength $B$ of the measured weak magnetic field. Parameters are: $J=0,~\Delta_a=0$, and other parameters are the same as in
		Fig.~2. }
\end{figure}

\begin{figure*}[t]
	\centering
	\includegraphics[width=0.3\linewidth]{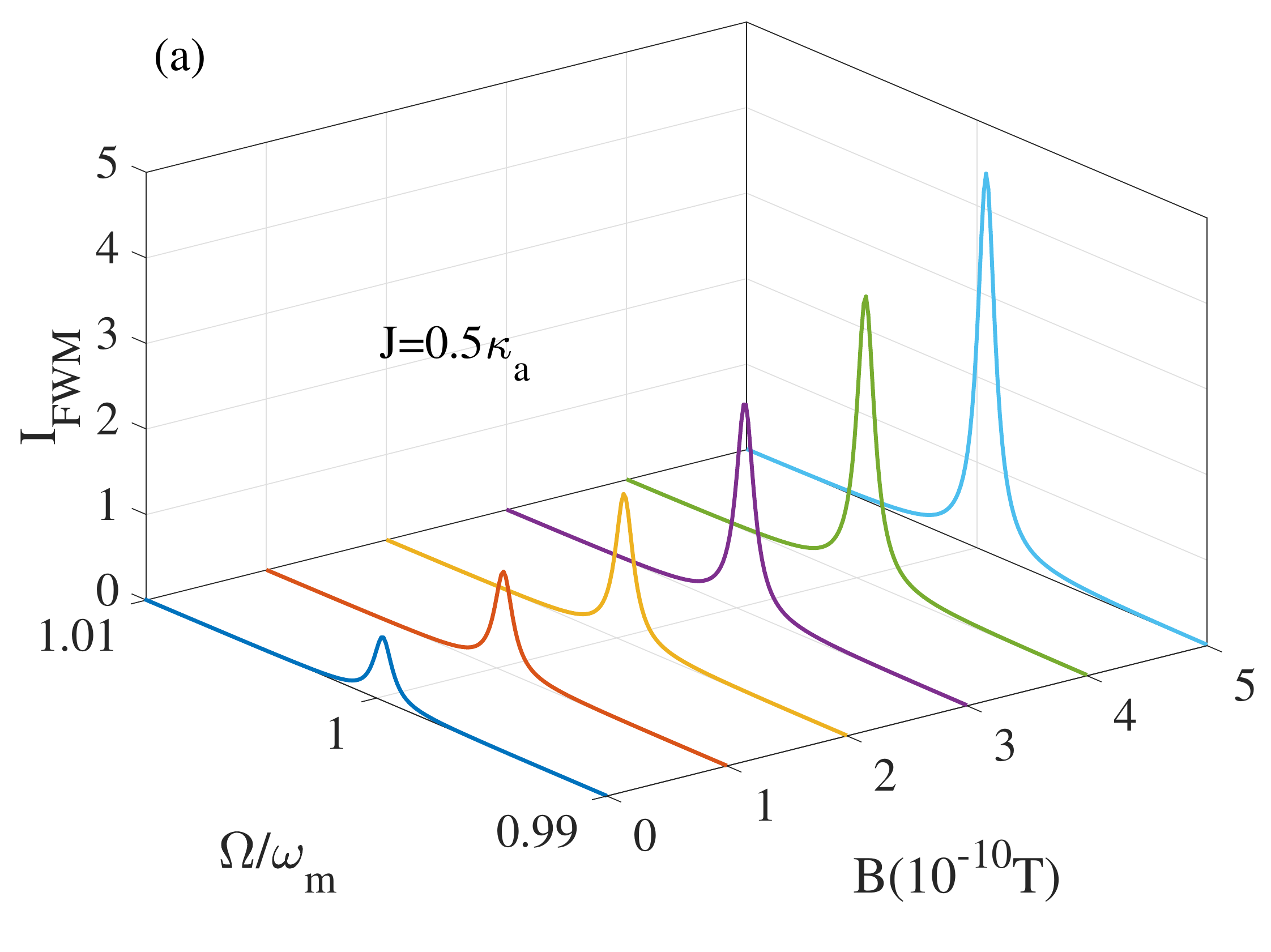}
	\includegraphics[width=0.3\linewidth]{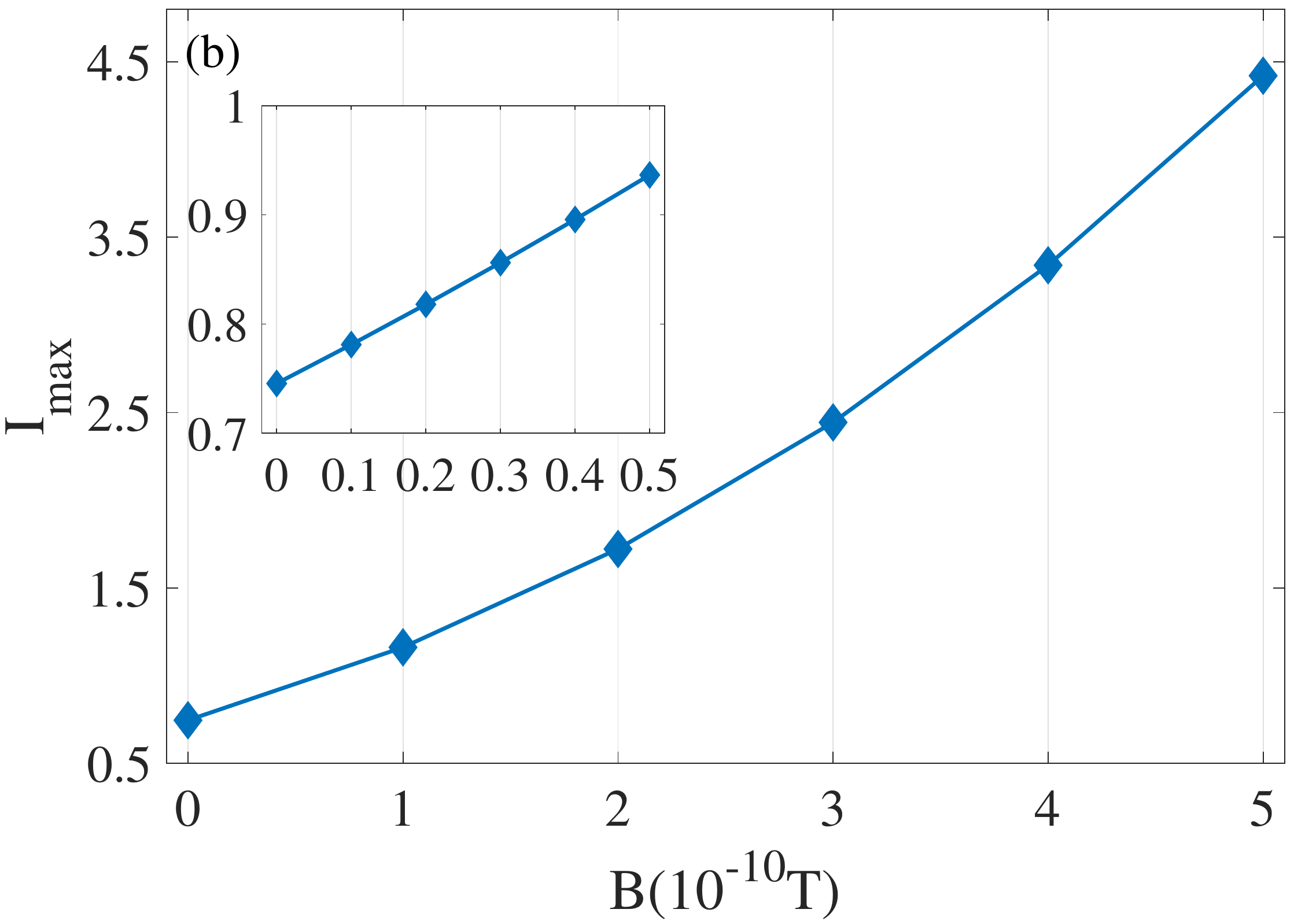}
	\includegraphics[width=0.3\linewidth]{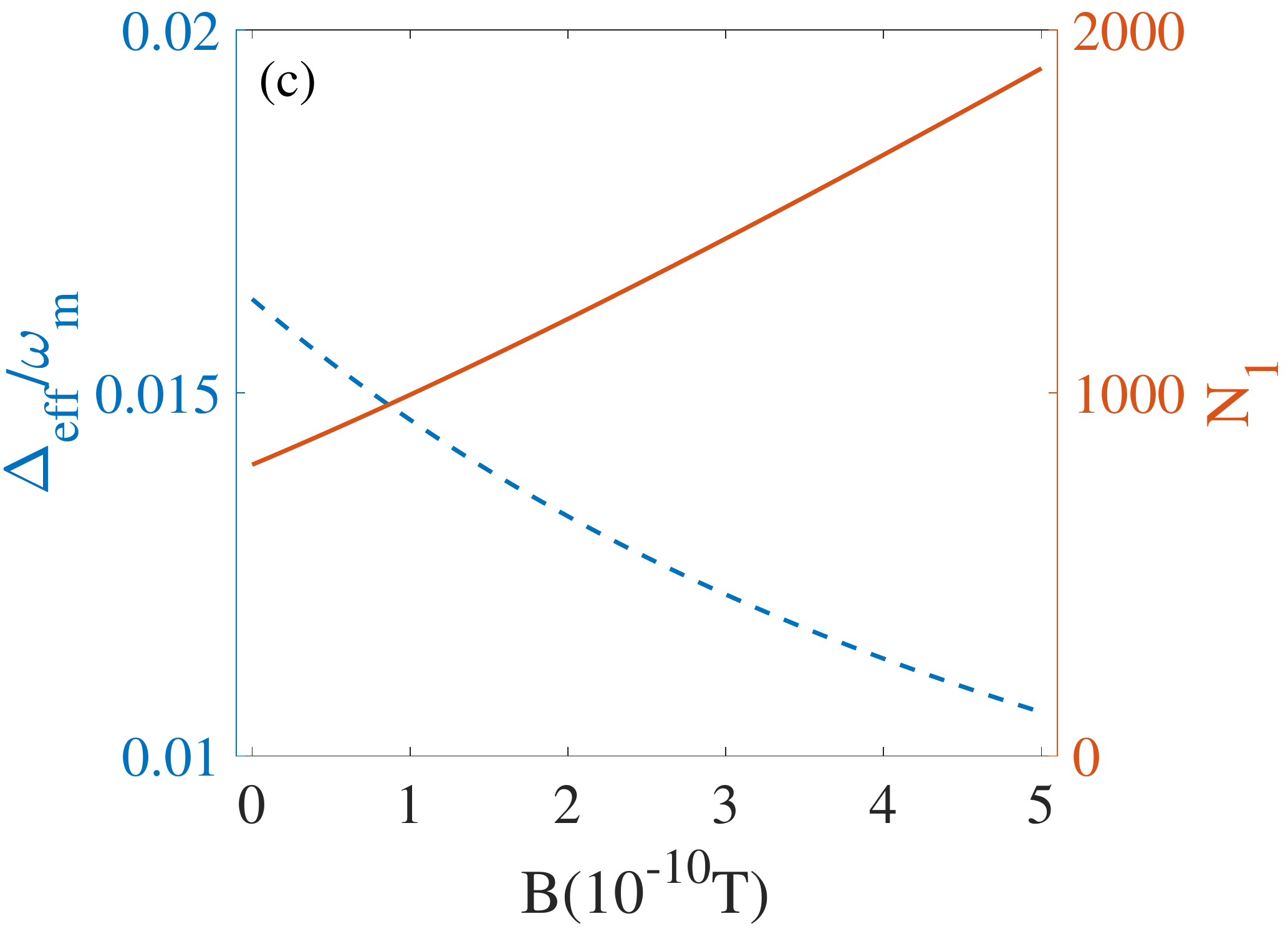}
	\caption{(color online) (a) FWM intensity spectrum as a function of detuning for different measured magnetic field. 
		(b) Maximum value $\textrm{I}_{\textrm{max}}$ of the FWM intensity as a function of magnetic field strength. 
		(c) Effective detuning $\Delta_{\textrm{eff}}$ and average photon number $N_1$ of loss-cavity as a function of
		magnetic field strength. Parameters are: $J=0.5\kappa_{a},~\Delta_{a}=0$, and other parameters are the same as in Fig. 2.}
\end{figure*}
$\textit{Magnetic field sensing with $\mathcal{PT}$-symmetric COMS.}$---\\Now, we study the weak magnetic field sensing based on the
$\mathcal{PT}$-symmetric COMS. As shown in Fig.~2(a), one can find that the FWM intensity generated with the $\mathcal{PT}$-symmetric
COMS ($J=0.5\kappa_a$) can be enhanced by two orders of magnitude compared to the resonantly driven single loss COMS ($J=0$), which
will be easier to observe in experiments. Besides, as shown in Fig.~4(a), the FWM intensity spectrum is plotted as a function of the detuning for
different magnetic field strengths. From the cures, we can see that the FWM intensity increases with the increase of the magnetic field, which is opposite to the
change trend of the resonantly driven single loss COMS. As expected, the $\mathcal{PT}$-symmetric COMS shows a higher sensitivity 
for the change of the weak magnetic field
compared to Fig.~2(c). In order to better show the dependence of the FWM intensity on the measured magnetic field, we plot the maximum value 
$\textrm{I}_{\textrm{max}}$ of
the FWM intensity spectrum as a function of the magnetic field strength, as shown in Fig.~4(b). From the inset of Fig.~4(b), one can see that the
$\mathcal{PT}$-symmetric COMS can even distinguish the magnetic field with strength 10$^{-11}$T based on the change of the FWM intensity, which is
increased by two orders of magnitude compared to the single loss COMS. Similarly, this enhanced sensitivity can be understood as follows: for
the $\mathcal{PT}$-symmetry-breaking COMS, the effective decay rate of loss-cavity can be completely balanced by the optical gain of 
gain-cavity, i.e., $\kappa_{\textrm{eff}}=0$ [see Eq.~(9)]. Meanwhile, as shown in Fig.~4(c), the effective detuning of loss-cavity can be close to the resonant
condition with the increase of the measured magnetic field, which also can be seen from the increasing trend of the average photon number with the
increase of the magnetic field.

\begin{figure*}
	\centering
	\includegraphics[width=0.3\linewidth]{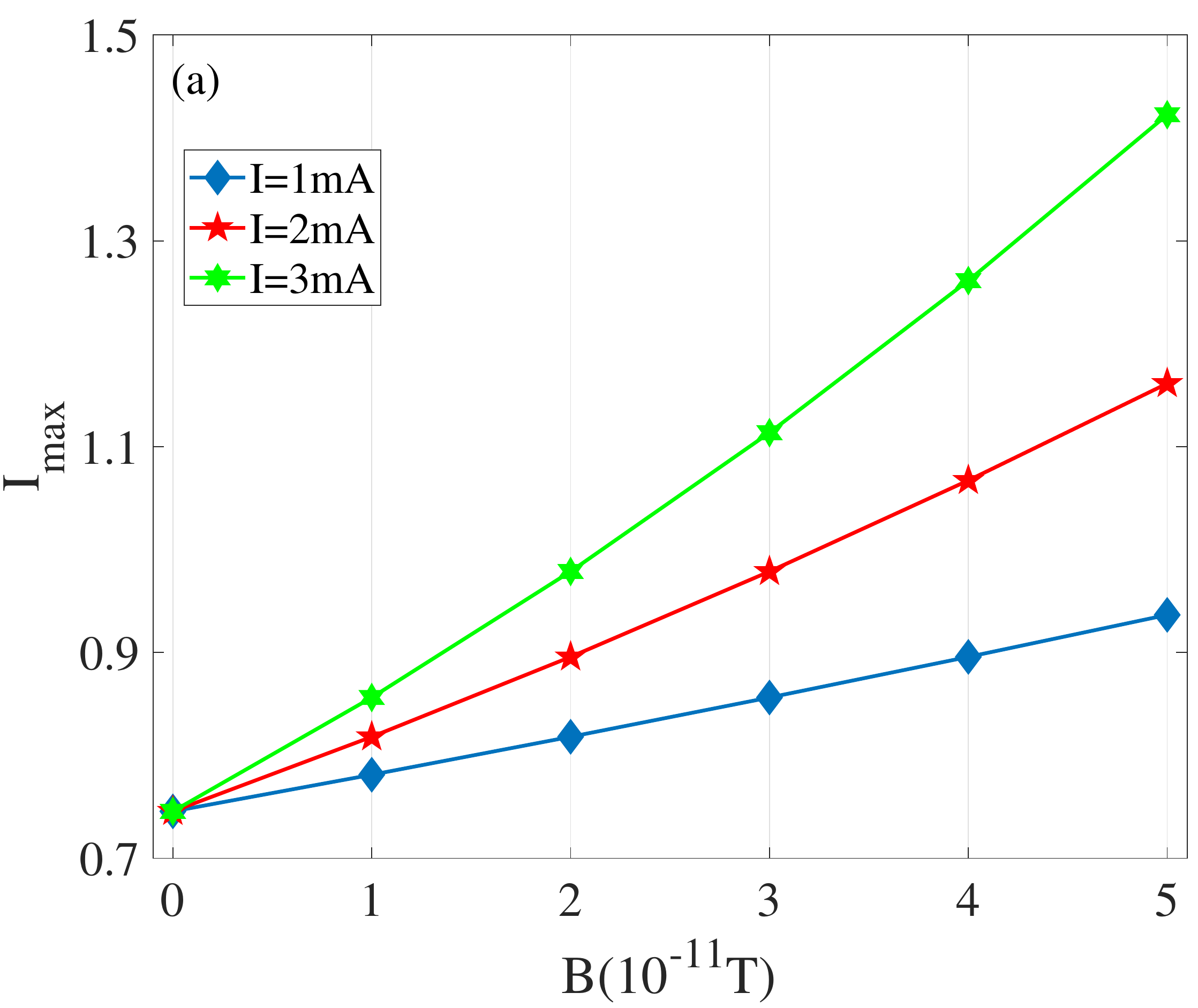}
	\includegraphics[width=0.3\linewidth]{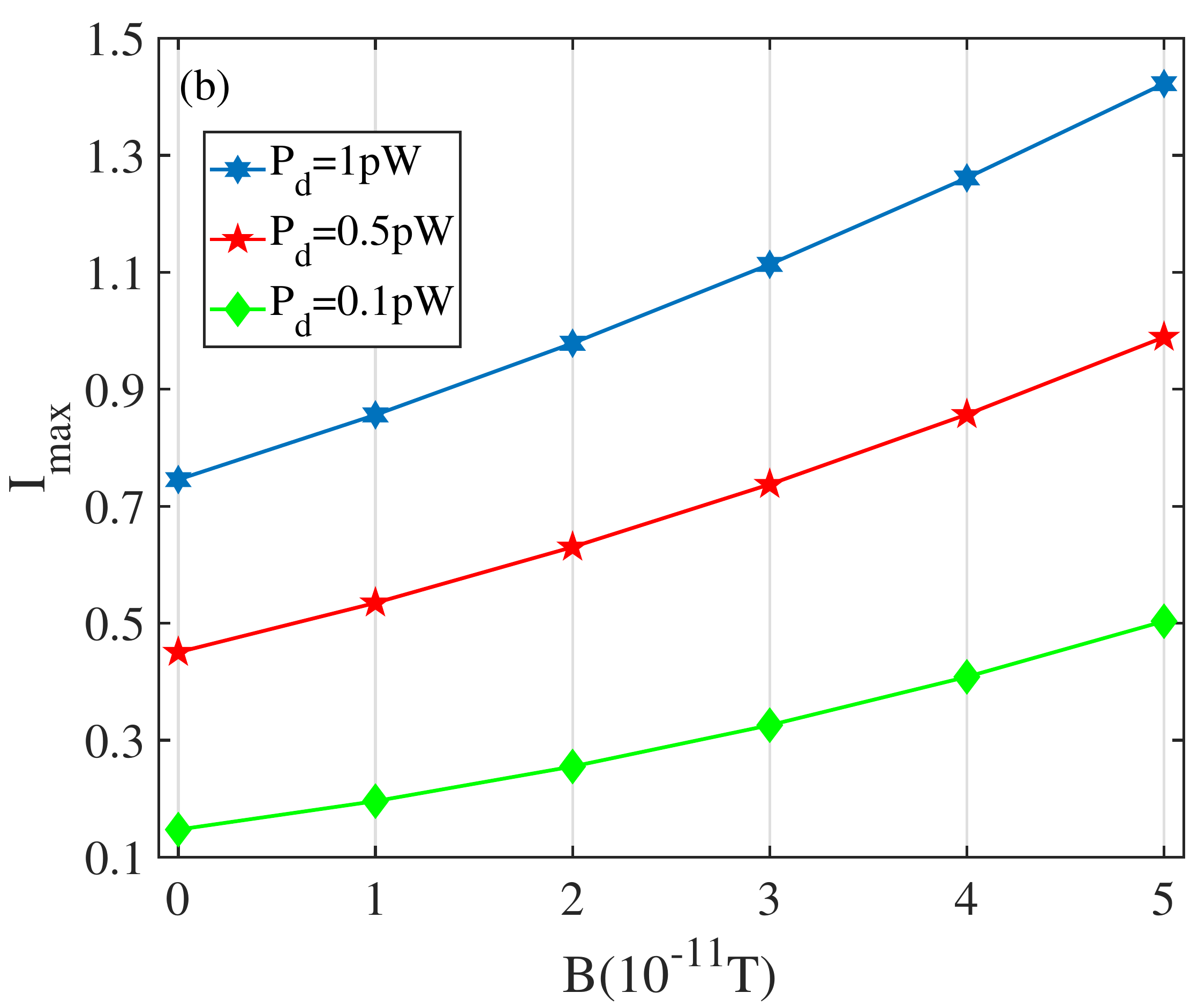}
	\includegraphics[width=0.3\linewidth]{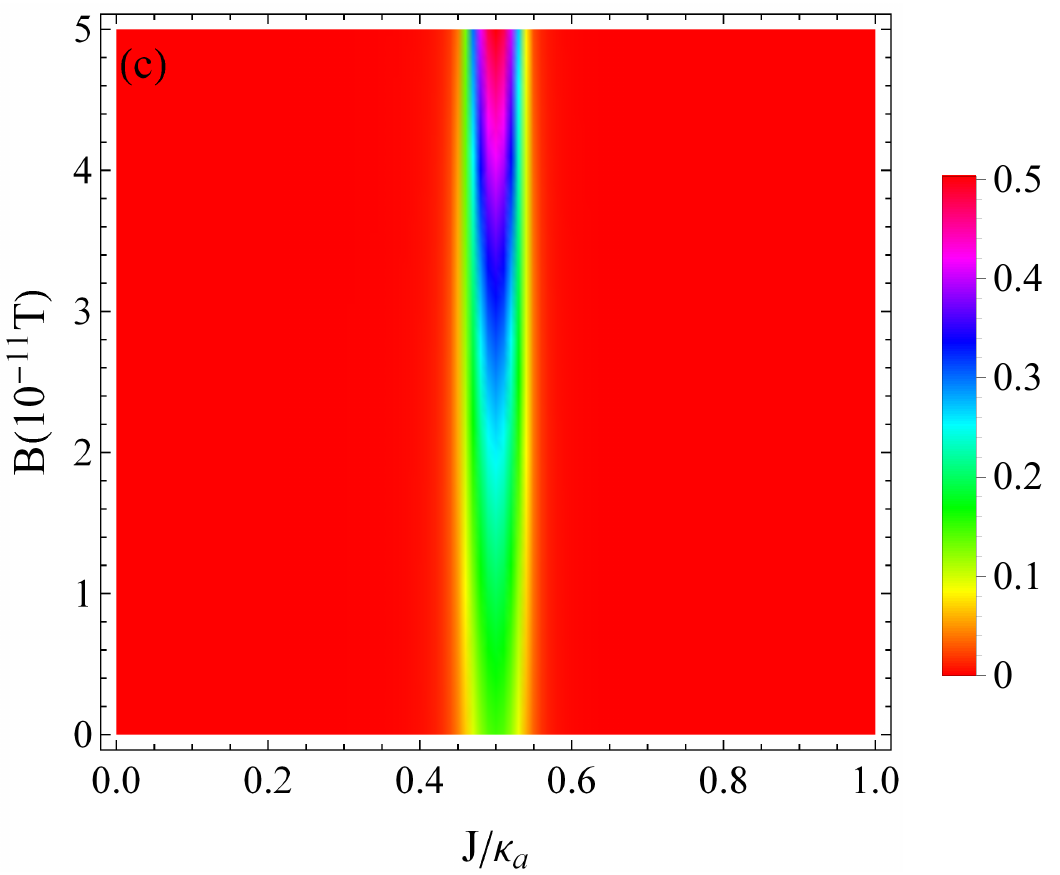}
	\caption{(color online). Maximum value of FWM intensity as a function of magnetic field strength for (a) different current intensities
		and (b) different driving powers. (c) FWM intensity spectrum as a function of magnetic field strength and tunneling coupling.
		Parameters are: (a) $P_{d}=1\textrm{pW},~I=1\textrm{mA},~2\textrm{mA}$ and $3\textrm{mA}$; (b) $I=3\textrm{mA},~P_{d}=1\textrm{pW},~0.5\textrm{pW}$ and $0.1\textrm{pW}$; (c) $I=3\textrm{mA},~P_{d}=0.1\textrm{pW}$, and other parameters are the	same as in Fig.~4.}
\end{figure*}
$\textit{Improvement\! of\! sensitivity\! of\! magnetic field sensing.}$---\\Before improving the sensitivity of the measured weak magnetic field, we should define a
sensitivity coefficient $ \eta$ to quantify the sensing performance. To distinguish different intensities generated by the FWM process, it is convenient to use
contrast as the sensitivity coefficient, i.e.,
\begin{equation}
\eta=\frac{\textrm{I}_{\textrm{max}}^{\textrm{n+1}}-\textrm{I}_{\textrm{max}}^{\textrm{n}}}{\textrm{I}_{\textrm{max}}^{\textrm{n+1}}+\textrm{I}_{\textrm{max}}^{\textrm{n}}}\times 100\%,~~(\textrm{n}\geq1)
\end{equation}  
with $\textrm{n}$ representing different FWM intensities. The larger the value of $\eta$, the easier for us to distinguish the two different FWM intensities.

In our $\mathcal{PT}$-symmetric COMS, the oscillator is subjected to optomechanical and electromagnetic interactions simultaneously. In order to improve
the sensitivity of magnetic field sensing, we can first increase the interaction between oscillator and magnetic field by increasing the current intensity. As
shown in Fig.~5(a), with the increase of current intensity, each FWM intensity corresponding to the same magnetic field strength can be enhanced
dramatically.  On the other hand, we can also decrease the driving power of loss-cavity, which can prevent the electromagnetic interaction from being
drowned in the optomechanical interaction.  Meanwhile, when the $\mathcal{PT}$-symmetry-breaking phase occurs, the effective decay rate of
loss-cavity is just zero, which can avoid the effects of decay rate on the system under the weak driving condition. Hence, decreasing the driving power can
better reflect the enhancement effect of the $\mathcal{PT}$-symmetry breaking on the magnetic field sensing of system. This is also reflected in
Ref.~\cite{lu2015p}, where a
$\mathcal{PT}$-symmetry-breaking chaos in optomechanics was realized with a ultralow driving threshold ($P_{d}=0.02\textrm{pW}$). As shown in Fig.~5(b), with the
decrease of driving power, although the FWM intensity decreases, the contrast of each $10^{-11}$T of measured magnetic field increases dramatically. For
example, based on the definition of contrast, the average contrast of each $10^{-11}T$ can be increased from $6.45\%,~7.85\%$ to $12.2\%$ for the
corresponding driving power with $\textrm{1pW,~0.5pW}$
and $\textrm{0.1pW}$. Moreover, we also analyze the effect of the tunneling coupling between gain- and loss-cavity on the magnetic field sensing, as shown in 
Fig.~5(c). One can find that in the $\mathcal{PT}$-symmetry-breaking phase (i.e., $J=0.5\kappa_{a}$), the contrast for the measured weak magnetic field is larger
than other parameter regions, which also shows that the $\mathcal{PT}$-symmetry breaking can enhance the performance of magnetic field sensing.

$\textit{Conclusions.}$---We have investigated the performance of the $\mathcal{PT}$-symmetry-breaking enhanced cavity optomechanical magnetometer based
on the nonlinear FWM process and analyzed the improvement of sensitivity for the magnetic field sensing. We showed that when
the $\mathcal{PT}$-symmetry-breaking phase occurs, the FWM intensity can be enhanced by two orders of magnitude compared to the conventional single
loss COMS, meanwhile, the measurement precision can also be increased from $10^{-9}$T to $10^{-11}$T. Our work uses the combined effects between
$\mathcal{PT}$-symmetry and motional sideband to enhance the performance of cavity optomechanical magnetometer, which is a significant improvement for
the magnetometer based on COMS and will have wide applications in the quantum sensing.

This work was supported by the National Key Research and Development Program of China (Grants No.~2017YFA0304202 and No.~2017YFA0205700), the
NSFC (Grants No.~11875231 and No.~11935012), and the Fundamental Research Funds for the Central Universities through Grant No.~2018FZA3005.

%%%%%%%%%%%%%%%%%%%%%%% References %%%%%%%%%%%%%%%%%%%%%%%%%
\bibliographystyle{apsrev4-2}
\bibliography{references}
\end{document}